\documentclass[aps,prl,superscriptaddress,twocolumn,floatfix]{revtex4-2}

\usepackage{color} 
\usepackage{hyperref}
\usepackage{amsmath,amssymb,mathrsfs}
\usepackage{psfrag}
\usepackage{graphicx}
\usepackage{graphics}
\usepackage{float}
\usepackage{subfigure}
\usepackage{epsfig}
\usepackage{bm}
\usepackage{verbatim,color}
\usepackage{braket}
\usepackage{import}

\begin{document}

\title{Phenomenological model of decaying Bose polarons}

\author{R. Alhyder}%
\email{Corresponding author: ragheed.alhyder@ist.ac.at}
\affiliation{Institute of Science and Technology Austria
(ISTA), Am Campus 1, 3400 Klosterneuburg, Austria
}%
\affiliation{Department of Physics and Astronomy, Aarhus University, Ny Munkegade 120, DK-8000 Aarhus C, Denmark
}%
\author{G. M. Bruun}%
\affiliation{Department of Physics and Astronomy, Aarhus University, Ny Munkegade 120, DK-8000 Aarhus C, Denmark
}%

\author{T. Pohl}%
\affiliation{Institute for Theoretical Physics, Vienna University of Technology,
Wiedner Hauptstraße 8-10, 1040 Vienna, Austria
}%
\affiliation{Department of Physics and Astronomy, Aarhus University, Ny Munkegade 120, DK-8000 Aarhus C, Denmark
}%

\author{M. Lemeshko}
\affiliation{Institute of Science and Technology Austria
(ISTA), Am Campus 1, 3400 Klosterneuburg, Austria
}%

\author{A. G. Volosniev}
\affiliation{Department of Physics and Astronomy, Aarhus University, Ny Munkegade 120, DK-8000 Aarhus C, Denmark
}%
\affiliation{Institute of Science and Technology Austria
(ISTA), Am Campus 1, 3400 Klosterneuburg, Austria
}

\begin{abstract}
Cold atom experiments  show that a mobile  impurity particle immersed in a weakly interacting Bose-Einstein condensate  forms a 
well-defined quasiparticle (Bose polaron) for weak to moderate impurity-boson interaction strengths, whereas 
a significant line broadening is consistently observed for strong interactions.
 Motivated by this, we introduce a  
phenomenological theory based on the assumption that the most  relevant states are characterized by the impurity 
correlated with at most one boson, since they have the largest overlap with the uncorrelated states to which the most common experimental probes 
couple. These experimentally relevant states can however decay to lower energy states characterised by correlations involving multiple bosons, and we model this  
using a minimal variational wave function combined with a complex impurity-boson interaction strength. We first motivate this approach by 
comparing to a more elaborate theory that includes correlations with up to two bosons. Our phenomenological  model is shown to recover the main results of two recent experiments probing both the spectral and the non-equilibrium  properties of the 
Bose polaron. Our work offers an intuitive framework for analyzing experimental data and highlights the importance of understanding 
 the complicated problem of the Bose polaron decay in a many-body setting. 
\end{abstract}

\maketitle

\textit{Introduction.}
Polaron quasi-particles are key to our description of mobile particles interacting with a surrounding many-body medium. 
Quantum simulation experiments with cold atoms have provided a wealth of detailed information advancing our 
understanding of  such polarons across a wide range of parameter regimes and interaction 
strengths, both regarding their spectral~\cite{schirotzekObservationFermiPolarons2009, nascimbeneCollectiveOscillationsImbalanced2009, kohstallMetastabilityCoherenceRepulsive2012, Koschorreck2012, Scazza2017, DarkwahOppong2019, ness2020observation, Fritsche2021, Fukuhara2013, hu2016bose, jorgensen2016, Yan2020, Baroni2024,massignan2014polarons}
and non-equilibrium properties~\cite{cetina2015,Cetina2016,skouNonequilibriumQuantumDynamics2021a,Skou2022,morgenQuantumBeatSpectroscopy2023,etrych2024universal,Grusdt2024review}. This in turn, reinforced by a wealth of theoretical advancements on two-dimensional polaron physics \cite{Casteels2012,Grusdt2016,Hryhorchak2020,Ardila2020,Isaule2021,CrdenasCastillo2022,Alhyder2022,Nakano2024}, has been used for interpretation of new experiments that create polarons in two-dimensional semiconductors, demonstrating the universal importance of this problem~\cite{massignan2025polaronsatomicgasestwodimensional}.

In spite of these advances, experimental data indicate that our understanding of impurity particles  in Bose-Einstein condensates (BECs) 
is incomplete. In particular, the observed spectral `polaronic' signal is always broad for strong interactions,
even for recent experiments performed in box potentials where the density is homogeneous apart from at the edges, 
which should reduce any density broadening~\cite{etrych2024universal}. 
The consistent observation of a broad spectral peak rather than a sharp quasiparticle peak naturally raises questions about the spectral weight and even existence of a long lived  Bose polaron for strong interactions. 
Second, the non-equilibrium impurity dynamics for repulsive interactions were recently observed to lose coherence much more rapidly than expected~\cite{morgenQuantumBeatSpectroscopy2023,etrych2024universal}. 

A major reason for our incomplete understanding of the Bose polaron is that contrary to the Fermi polaron, there is no Pauli exclusion principle to suppress correlations between the impurity and $N\ge 2$ bosons including 
the possible formation of  bound states (trimers, tetramers, $\ldots$)~\cite{braaten2006universality}. 

On the one hand, such few-body correlations in a many-body setting make the theoretical description of Bose polarons  challenging~\cite{massignan2025polaronsatomicgasestwodimensional}. On the other hand, experiment shows that 
relatively simple theories such as the ladder approximation~\cite{rath2013} including only 2-body correlations are able to recover
the peak position (but not the width) of the  spectral signal fairly well, as well as the 
non-equilibrium formation dynamics except the damped beating signal between the repulsive and attractive polaron~\cite{morgenQuantumBeatSpectroscopy2023}.

Motivated by this, we explore here the conjecture that  
the experimentally important polaron states  are characterized by the impurity being 
dressed by at most one boson. 
These many-body states do not necessarily correspond to the ground state but rather to broad resonances embedded in a many-body continuum. In other words, these states have a finite lifetime due to decay to low-lying states with a small spectral weight. 

We first frame this conjecture  by analyzing the spectral properties of a mobile impurity in a BEC using a well-known variational wave function that includes up to two Bogoliubov modes. This analysis suggests that the polaron-like state 
where the impurity is dressed by one boson appears 
as a broad resonance with a large spectral weight. This state lies above the long lived state with a lower energy but also a small spectral weight.
Instead of microscopically calculating the decay rate of the polaron state, which is very challenging 
as it requires the inclusion of all relevant 
decay channels in a many-body system, we develop a 
phenomenological extension of the ladder approximation that describes the dominant spectral response. 
We then show that this phenomenological theory captures both the spectral and non-equilibrium features observed in two recent experiments~\cite{etrych2024universal,morgenQuantumBeatSpectroscopy2023}.

\begin{figure}[t]
	\centering
\includegraphics[width=1\columnwidth]{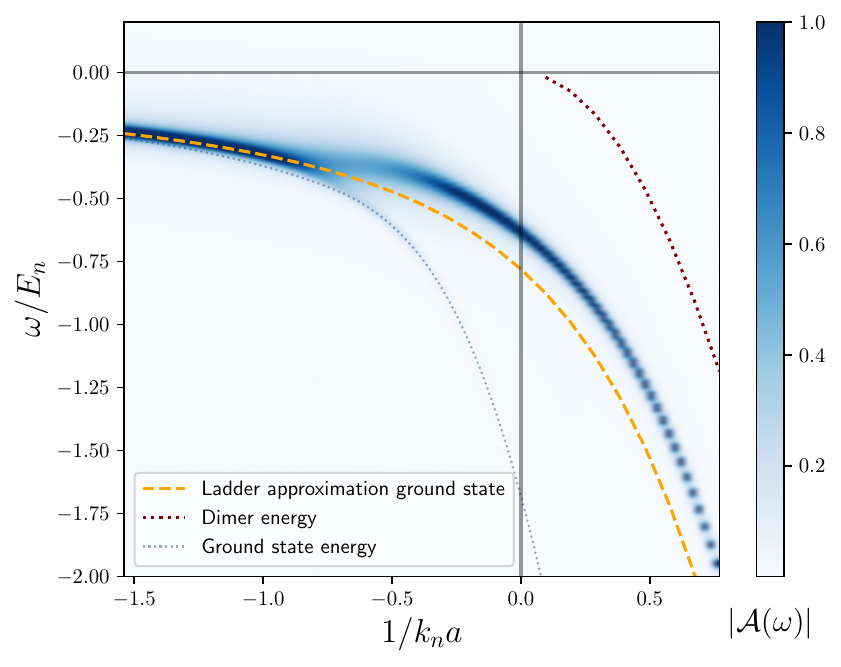}
	\caption[short]{ The zero momentum impurity spectral function 
    calculated using the ansatz in Eq.~(\ref{eq:ansatz}) for $n_0^{1/3}a_- = -1$ and $n_0^{1/3}a_{\mathrm{B}} = 0.02$. 
    The ground state energy is shown as a faint dotted curve,
        the dashed yellow curve shows the ground state energy within the ladder approximation, and
        the dotted curve shows the energy of the dimer, $-1/a^2$. 
   We have defined  $E_n = (6\pi^2n_0)^{2/3}/2$  and $1/k_n=1/\sqrt{2 E_n}$. Note that the spectral function is computed numerically for a discrete set of interaction strengths, which accounts for the apparently discontinuous behavior of the high‑energy branch in the repulsive regime.
   }
\label{fig:spectral_function}
\end{figure}

{\it Framework}. 
We consider an impurity particle of mass $m$ immersed in a homogeneous weakly interacting BEC with density $n_0$.
Since our focus is on the recent $^{39}$K experiments~\cite{etrych2024universal,morgenQuantumBeatSpectroscopy2023}, we assume that the mass of the bosons is also $m$, but
the theory can be  easily extended to mass-imbalanced systems. We use the two-channel Hamiltonian $\hat H=\hat H_0+\hat H_\text{int}$ 
to describe the system with 
\begin{align}
	\hat{H}_0 &= \sum_{\mathbf{k}} 
	\epsilon_{\mathbf k}
	\hat{c}^\dagger_{\mathbf{k}} 
	\hat{c}_{\mathbf{k}}
	+\sum_{\mathbf{k}} 
	\xi_{\mathbf{k}} 
	\hat{\beta}^\dagger_{\mathbf{k}} 
	\hat{\beta}_{\mathbf{k}}
	+\sum_{\mathbf{k}} 
	(k^2/4 + E_0)
	\hat{d}^\dagger_{\mathbf{k}} 
	\hat{d}_{\mathbf{k}}\nonumber
\\
      \hat{H}_\text{int} &=
	g\sum_{\mathbf{k}}(\sqrt{n_0}\hat{d}^\dagger_{\mathbf{k}} \hat{c}_{\mathbf{k}} 	 
 +\sum_{\mathbf{q}}
 \hat{d}^\dagger_{\mathbf{q}} \hat{c}_{\mathbf{q}-\mathbf{k}} \hat{b}_{\mathbf{k}}
	  )+\text{h.c.},
      \raisetag{24pt}
    \label{eq:two-channel-Hamiltonian}
\end{align}
where the operators $\hat{c}_{\mathbf k}^\dagger/\hat{\beta}^\dagger_{\mathbf k}/\hat{b}^\dagger_{\mathbf k}$ create an impurity/Bogoliubov mode (phonon)/boson 
with momentum $\mathbf k$, and $\hat{d}_{\mathbf k}^\dagger$ creates a closed channel (bare Feshbach) molecule, a two-body bound state of the impurity with one bath boson. 
We have as usual 
$\hat{b}_{\mathbf{k}} = u_{\mathbf{k}}\hat{\beta}_{\mathbf{k}}-v_{\mathbf{k}}\hat{\beta}^\dagger_{-\mathbf{k}}$ with $
u_{\mathbf{k}} = \sqrt{(\epsilon_{\mathbf k}+n_0 g_\mathrm{B})/(2\xi_{\mathbf k})+1/2}$, 
    $v_{\mathbf{k}} = \sqrt{(\epsilon_{\mathbf k}+n_0 g_\mathrm{B})/(2\xi_{\mathbf k})-1/2},$  and   $ g_\mathrm{B}=4\pi a_\mathrm{B}$ with $a_\mathrm{B}>0$ being the boson-boson  scattering length.
The energies are $\epsilon_{\mathbf k}=k^2/2$ and
$\xi_{\mathbf k}=\sqrt{\epsilon_{\mathbf k}(\epsilon_{\mathbf k}+2g_Bn_0)}$. Here and in what follows we take the system volume, $\hbar$ and $m$ to be unity.
The bare molecule energy $E_0$, open-closed channel coupling $g$, and ultraviolet momentum 
cut-off $\Lambda$ can be eliminated in favour of the  boson-impurity scattering length $a$ and effective range $R_e$~\cite{bruun2004,Gogolin2008}, which also gives rise to  
the three-body parameter $a_- = 2467 \, R_e$ determining when the first Efimov trimer becomes stable~\cite{yoshida2018}; see the Supplemental Material below.

To find the eigenstates of the Hamiltonian in Eq.~(\ref{eq:two-channel-Hamiltonian}), we use the variational ansatz~\cite{lisarma2014,levinsen2015impurity} 
\begin{align}
	\begin{split}
	&\ket{\Psi} = \Big( A \hat{c}^\dagger_0  + \sum_{\mathbf{k}} B_{\mathbf{k}} \hat{c}^\dagger_{-\mathbf{k}} \hat{\beta}^\dagger_{\mathbf{k}} +  \frac{1}{2} \sum_{\mathbf{k}\mathbf{p}} C_{\mathbf{k},\mathbf{p}} \hat{c}^\dagger_{-\mathbf{k}- \mathbf{p}} \hat{\beta}^\dagger_{\mathbf{k}} \hat{\beta}^\dagger_{\mathbf{p}} 
	\\&+ D \hat{d}^\dagger_{0} +
	\sum_{\mathbf{k}} E_{\mathbf{k}} \hat{d}^\dagger_{-\mathbf{k}} \hat{\beta}^\dagger_{\mathbf{k}} 
	\Big) \ket{\text{BEC}}.
	\label{eq:ansatz}
    \raisetag{25pt}
	\end{split}
\end{align}
The first three terms  describe the dressing of the impurity by $N=0,1,2$ Bogoliubov modes whereas the fourth and the fifth terms give the closed channel molecule components of the wave function including the dressing by up to $N=1$ Bogoliubov mode. 
Such a wave function has been used to successfully recover the experimental results 
regarding the spectral maximum of an impurity in a BEC~\cite{massignan2025polaronsatomicgasestwodimensional,Grusdt2024review}.
Here, 
we will also use the ansatz in Eq.~(\ref{eq:ansatz}) with time-dependent coefficients $A(t)$, $B_{\mathbf{k}}(t)$, $C_{\mathbf{k},\mathbf{p}}(t)$, $D(t)$, and $E_{\mathbf{k}}(t)$ to study quench dynamics; see the Supplemental Material below.

For the ground state, the variational parameters in Eq.~(\ref{eq:ansatz}) can  be found by minimizing the expectation value  $\langle\Psi|\hat{H}|\Psi\rangle$ of the Hamiltonian. 
To overcome the challenge of accessing the excited states, we use the 
 Krylov subspace method~\cite{Yousef2003,Yousef2011,liesen2012krylov,nandy2024quantumdynamicskrylovspace}, which 
 iteratively constructs a reduced basis that captures the action of the Hamiltonian on the initial state, thereby forming a Krylov subspace. The implementation details are given in the Supplemental Material below. It offers efficient access to the low-energy spectrum, essential for the present study. To construct the spectral function from the spectrum, we apply the standard finite‑broadening prescription; that is, the delta peaks associated with the discrete eigenspectrum are replaced by narrow Lorentzian functions.

{\it Spectral function.}  The zero momentum impurity spectral function obtained from the Krylov subspace method is shown in Fig.~\ref{fig:spectral_function} for the parameters $n_0^{1/3}a_- = -1$ and $n_0^{1/3}a_{\mathrm{B}} = 0.02$  allowing for a direct comparison
with Ref.~\cite{levinsen2015impurity}.  Near the unitary limit $1/k_n a=0$, the spectral function features two branches:
A sharp low-energy branch,  which is barely visible, reflecting its vanishing spectral weight, in agreement with the properties ground state of the problem studied in Ref.~\cite{levinsen2015impurity}, 
and a broad high-energy branch present at strong interactions. 
We furthermore see that the maximum in the spectral density of this branch is in the vicinity of the results of 
the ladder approximation, which is equivalent to including terms with maximally one Bogoliubov mode in Eq.~\eqref{eq:ansatz}. 
This indicates that the upper branch  corresponds to states dressed by one boson, which decay to the  ground state dressed by two bosons within this approximation. 
Importantly, the spectral weight of the upper branch vastly exceeds that of the ground state branch  
suggesting that it is the primary contributor to most experimental observables.
We find that these qualitative features are robust for  $n_0^{1/3}|a_-|\gtrsim 1$. 

{\it Phenomenological model.---} 
The ansatz in Eq.~\eqref{eq:ansatz} has been extended to include up to three Bogoliubov modes for $a_\mathrm{B}=0$ and $1/a$  
obtaining no clear convergence for the ground state energy~\cite{yoshida2018}, which may reflect an underlying bosonic orthogonality catastrophe for 
an impurity in an ideal BEC for a single channel interaction~\cite{Guenther2021}. 
In general, calculating the full spectral response from both ground and excited states encounters scalability challenges with the number of included Bogoliubov modes. Furthermore, the Hamiltonian Eq.~\eqref{eq:two-channel-Hamiltonian} neglects boson-boson interactions beyond mean-field, which makes the accuracy of this approach unclear for 
states involving the dressing by multiple bosons~\cite{Levinsen2021,Christianen2024,Schmidt2022,massignan2025polaronsatomicgasestwodimensional}. 
We therefore adapt in  the following a simpler phenomenological approach.

Instead of the multichannel model given by Eq.~\eqref{eq:two-channel-Hamiltonian}, we now use the interaction 
\begin{equation}
	\hat {\mathcal{H}}_\text{int}=
	g_I\sum_{\mathbf {k'kq}}c_{\mathbf k+\mathbf q}^\dagger c_{\mathbf k}b_{\mathbf k'-\mathbf q}^\dagger b_{\mathbf k'}.
    \label{eq:model_complex}
\end{equation}
For real values of the interaction strength $g_I$, Eq.~\eqref{eq:model_complex} is simply the standard single-channel point interaction,
but we will instead employ a \emph{complex} $g_I$  where 
the imaginary part
introduces an imaginary part to the energy~\cite{yamamotoTheoryNonHermitianFermionic2019, bouchouleEffectAtomLosses2020, yamamotoCollectiveExcitationsNonequilibrium2021, bouchouleBreakdownTanRelation2021, wangComplexContactInteraction2022}. 
This allows us to describe the  decay and decoherence of the polaron state due to any omitted part of the Hilbert space in a relatively straightforward way. 
An imaginary part of $g_I$ can in fact be shown to naturally emerge due to $N\ge 3$-body losses, as discussed in the Supplemental Material below, but we will in the following not provide a microscopic analysis of the origin of this loss, regarding it instead as a free parameter.

Moreover, since we know that the experimentally observed maximum of the spectral response is fairly accurately recovered by the ladder approximation~\cite{massignan2025polaronsatomicgasestwodimensional}, which is equivalent to a variational wave function 
truncated at a single Bogoliubov mode  level, we fix $C=D=E=0$ in Eq.~(\ref{eq:ansatz}).
As Fig.~\ref{fig:spectral_function} shows, the physical reason behind this approximation is that the states dressed with at most one boson  have a large  overlap (in comparison to low-lying states) with the non-interacting plane wave state to which both RF and Ramsey experiments couple. Consequently, the dominant experimental signal does not, in general, track the ground state but an excited, one-boson-dressed polaron that decays into low-lying states.

The calculated spectral function depends on $a_\mathrm{B}$ rather weakly in the ladder approximation~\cite{massignan2025polaronsatomicgasestwodimensional}, and to 
gain physical intuition we first consider an ideal BEC with $a_\mathrm{B}=0$, which reproduces key spectral features at $T=0$~\cite{Drescher2024}. In this case, the pole of the impurity propagator is found by solving 
\begin{equation}
\omega\left(\frac{1}{g_I}+\frac{\Lambda}{2\pi^2}-i\frac{\sqrt{\omega}}{4\pi}\right)=n_0.
\label{eq:energy_cubic}
\end{equation}
To fit the experimental data, we  use 
\begin{equation}
\frac{1}{g_I}\simeq \frac{1}{4\pi a}-\frac{\Lambda}{2\pi^2}+i\frac{\Gamma+\gamma n_0^{1/3} |a|}{4\pi |a|}.
\label{eq:g_I_final}
\end{equation}
The first two terms are the standard prescription giving an undamped 
attractive polaron with energies $\omega\rightarrow4\pi n a$ and $\omega\rightarrow -1/a^2$ for $a\rightarrow 0^-$ and $a\rightarrow 0^+$ respectively, and a 
repulsive polaron with energy $\omega\rightarrow4\pi n a$ for $a\rightarrow 0^+$. As discussed above, these energies  recover 
reasonably well  the maximum of the observed spectroscopic signal. 

The last term in Eq.~\eqref{eq:g_I_final} moves the pole into the complex plane thereby introducing damping. In the spirit of our phenomenological approach, 
this term is constructed to interpolate between the observed linewidth in the weakly and strongly interacting regimes
rather than being based on microscopic calculations. 
 For weak interactions  $a\to0$, we find from Eqs.~\eqref{eq:energy_cubic}-\eqref{eq:g_I_final} that the complex energy is   
$\omega\simeq 4\pi n a-i4\pi n |a|\Gamma$ for $\Gamma\ll1$. This  recovers the mean-field energy with a damping that scales linearly with the scattering length as recently observed for the Bose polaron in a box potential~\cite{etrych2024universal}. Likewise, 
a straightforward but more cumbersome analysis yields  
$\omega\simeq -(4\pi n)^{2/3}-2 i \gamma (4\pi)^{1/3} n^{2/3}/3$ for $1/a\to0$, which also agrees with the observed $n^{2/3}$
scaling  of the damping at unitarity~\cite{etrych2024universal}. 

\begin{figure}
	\centering
 \includegraphics[width=1\columnwidth ]{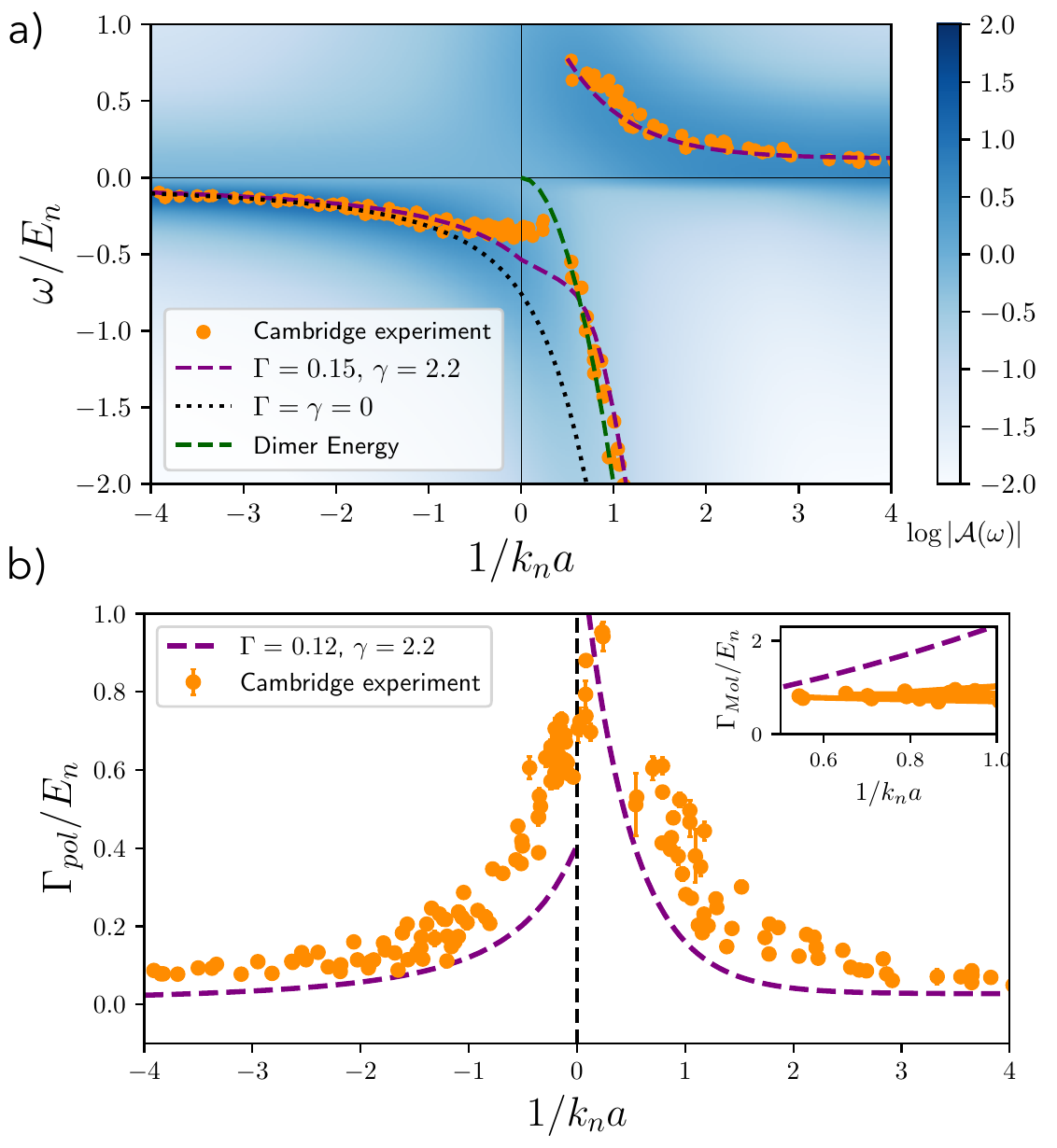}
	\caption[short]
{a) The zero momentum impurity spectral function from our phenomenological model. 
The black dotted curve shows the ladder approximation with real $g_I$,  the green dashed curve is the molecular state, and  the orange
dots are the experimental data of Ref.~\cite{etrych2024universal}.
b) Extracted widths $\Gamma_{pol}$ of the attractive polaron for $a<0$ and repulsive polaron for $a>0$. To match the experiment~\cite{etrych2024universal}, we use $n^{1/3} a_B = 0.005$, and for fitting the signal we use $\Gamma=0.12$ and
$\gamma=2.2$. Inset: Extracted width $\Gamma_{Mol}$ of the attractive polaron for $a>0$.}
 \label{fig:fig3}
\end{figure}

{\it Spectral properties.---} 
We now explore the spectral properties of a zero momentum impurity as predicted by 
our phenomenological model, focusing on comparing with the recent 
experiment~\cite{etrych2024universal}. Figure~\ref{fig:fig3}(a) 
plots the spectral function of the impurity obtained numerically using $\Gamma=0.12$ and
$\gamma=2.2$ with the experimentally realistic value  $n^{1/3}a_B=0.005$. We clearly see a broad spectral 
response with a large linewidth in the vicinity of the unitary limit.
Following Ref.~\cite{etrych2024universal}, we extract the position $E_p$ and  width  $\Gamma_{pol}$ of the signal, using a Lorentzian fit 
for the attractive polaronic branch ($E_p<0$) for $a<0$ and the repulsive branch ($E_p>0$) for $a>0$, and
Gaussian fit for the attractive branch ($E_p<0$) for $a>0$~\footnote{For each interaction value ($1/k_n a$) we analyze the zero-momentum impurity spectral function $A(\omega)$ on the relevant branch and fit a simple parametric line shape to a \emph{local} window around the peak.
When the branch is well described by a Lorentzian (attractive branch for $a<0$; repulsive branch for $a>0$), we fit
$$
L(\omega)=A\,\frac{\Gamma^2}{(\omega-\omega_0)^2+\Gamma^2}+C,
$$
and define the peak position and width as
$
E_p\equiv \omega_0,\, \Gamma_{\mathrm{pol}}\equiv \Gamma
$
(HWHM).
For the attractive branch at $a>0$, consistent with Ref.~[20], we use a Gaussian profile
$$
G(\omega)=A\,e^{-(\omega-\omega_0)^2/(2\sigma^2)}+C,
$$
and report
$
E_p\equiv \omega_0,\, 
\Gamma_{\mathrm{pol}}\equiv \sigma$.
Fits are performed with nonlinear least squares, and initial guesses are set by the local maximum of $A(\omega)$. }. The obtained widths 
are plotted in Fig.~\ref{fig:fig3}(b). We observe reasonable agreement with the experimental results of Ref.~\cite{etrych2024universal} plotted as yellow dots, both 
regarding the peak positions of the  branches \emph{and} their widths. A better agreement may be achieved by performing a global fit to determine the parameters $\gamma$ and $\Gamma$.

Both the experimental and theoretical data exhibit a shift of the attractive-branch maximum toward the molecular energy for $a>0$ as well as the asymmetric behavior of the widths of the attractive and 
repulsive branches, which is a manifestation of different physics at play. Indeed, the repulsive polaron 
 is inherently metastable even for $\gamma=\Gamma=0$ due to the presence of a low-lying molecular state~\cite{scazzaRepulsiveFermiBose2022}, therefore,  its width is weakly influenced by the complex interaction. It is worth noting that the agreement between the experimental and theoretical results presented in Fig.~\ref{fig:fig3} can be further improved by incorporating broadening mechanisms specific to this experiment, like magnetic field fluctuations, three-body losses and inhomogeneous density into our theoretical model. Including these mechanisms would increase the estimated polaron width and result in larger error bars for the peak position. Consistent with these trends, a recent solvable-model analysis reproduces central features of the observed spectra and provides a complementary microscopic perspective on the damping systematics~\cite{etrych2025fateimpuritystronglyinteracting}.

\begin{figure}
	\centering
\includegraphics[width=\columnwidth ]{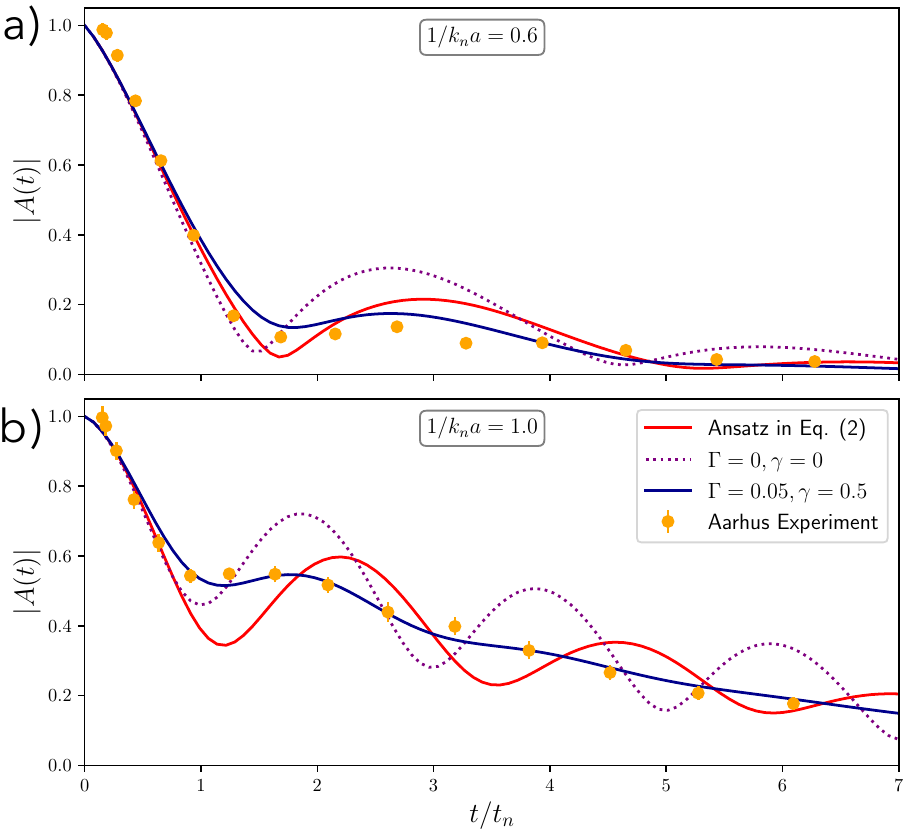}
	\caption[short]
{Time dynamics of the contrast $|A(t)|$ on the repulsive side of the Feshbach resonance for $(k_na)^{-1} = 0.6$ in panel $(a)$ and $(k_na)^{-1} = 1.0$ in $(b)$. Dotted purple curves correspond to the standard ladder approximation ($\gamma=\Gamma=0$); solid blue curves are calculated using our phenomenological model with $\Gamma=0.05, \gamma = 0.5$; red solid curves 
are from the ansatz Eq.\ \eqref{eq:ansatz} in the two-channel Hamiltonian~(\ref{eq:two-channel-Hamiltonian}) with $n^{1/3}a_- = -100, n^{1/3}a_B = 0.005$ similar to the experiment~\cite{morgenQuantumBeatSpectroscopy2023}. The orange dots represent experimental data~\cite{morgenQuantumBeatSpectroscopy2023}. We account for magnetic dephasing and three-body loss when computing the blue and red solid curves. 
}
 \label{fig:fig4}
\end{figure}

{\it Dynamical properties.---} 
The dynamical properties of impurities and polaron formation can be probed using Ramsey
interferometry, which measures the overlap 
$A(t)=\langle\text{BEC}|\hat c_{{\bf k}=0}(t)\hat c_{{\bf k}=0}^\dagger(0)|\text{BEC}\rangle$
between the state created by injecting a zero momentum impurity at time $t=0$ and taking it out again at 
time $t$~\cite{Goold2011,Knap2012,Schmidt_2018}. This technique was pioneered to explore Fermi polarons~\cite{Cetina2016,cetina2015} 
and subsequently modified to explore Bose polarons~\cite{skouNonequilibriumQuantumDynamics2021a}. 

Such experiments correspond to measuring $A(t)$ in Eq.~\eqref{eq:ansatz} with an 
 initial `bare' wave function for which $A(t=0)=1$ and $B_{\mathbf k}(t=0)=0$. 
The impurity is subsequently dressed by excitations in the BEC resulting in $|A(t)|<1$ and $B({\mathbf k},0)\neq0$~\footnote{It is easy to understand the time dynamics of $A(t)$ for a non-interacting Bose gas ($a_B=0$) using the following ansatz for $A(t)$, which fits the numerical data well; see the Supplemental Material below,
\begin{equation}
A(t)=\sqrt{Z}e^{-i \omega_1 t}+(1-\sqrt{Z})e^{-i\omega_2 t},
\label{eq:A_exact}
\end{equation}
where $|Z| = 8\pi n a\left(12\pi n a-\omega\right)^{-1}$ is the residue and $\omega_1$  ($\omega_2$) is the solution of  Eq.~(\ref{eq:energy_cubic}) with $\mathrm{Im}(\omega)<0$; in the limit $\Gamma=0$, $\omega_1$ ($\omega_2$) is the standard attractive (repulsive) polaron~\cite{Yan2020}. The coefficients $B$ read
\begin{equation}
B_\mathbf{k}(t)=\frac{\omega_1 \sqrt{Z}f_{\mathbf{k}}(\omega_1)}{(2\pi)^{3/2}(\omega_1-2\epsilon_{\mathbf{k}})}+\frac{\omega_2(1-\sqrt{Z})f_{\mathbf{k}}(\omega_2)}{(2\pi)^{3/2}(\omega_2-2\epsilon_{\mathbf{k}})},
\label{eq:B_exact}
\end{equation}
 where $f_{\mathbf{k}}(\omega)=e^{-i\omega t}-e^{-2i\epsilon_{\mathbf{k}} t}$.
The first term in $f_{\mathbf{k}}$ is due to the pole, and the second one is due to the bosonic continuum. The latter contribution describes the particles that move away from the impurity, and do not interact with it at later times. For an interacting Bose gas, this would correspond to sound waves emitted by the impurity (see, e.g.,  Ref.~\cite{Marchukov2021} for a further visualization of this process).}.
We compare the results with a recent experiment measuring $A(t)$ for $a>0$~\cite{morgenQuantumBeatSpectroscopy2023}. The results for $a<0$ are presented in the Supplemental Material below.

In Fig.~\ref{fig:fig4}, we present our calculations of the contrast $|A(t)|$ using different methods and compare them with the experimental
data~\cite{morgenQuantumBeatSpectroscopy2023}.  First, we see that the  ladder approximation  with a real interaction strength 
captures the initial decay of $|A(t)|$ for the present case just as well as for the attractive case $a<0$~\cite{skouNonequilibriumQuantumDynamics2021a}.
It also predicts  oscillations due to the interference between the repulsive and attractive polaron branches 
~\cite{morgenQuantumBeatSpectroscopy2023} (not present for $a<0$), which however are much more pronounced 
than what is observed experimentally even after trap averaging and magnetic field fluctuations are included~\cite{morgenQuantumBeatSpectroscopy2023}.
The results based upon the ansatz from Eq.~\eqref{eq:ansatz} are closer to the measured values featuring a quicker decay of $|A(t)|$.
 This demonstrates that a more accurate description of the attractive high-energy branch and its decay is important for describing the 
 observed dynamics~\footnote{Note that our phenomenological model agrees with the variational ansatz  Eq.~\eqref{eq:ansatz} regarding the dynamics when  
$\Gamma =0.01$ and $\gamma = 0.1$.}. Still, this improved description does not fully agree with the experimental results where the decay of oscillations is even more pronounced.

Figure \ref{fig:fig4} shows that our phenomenological model with $\Gamma= 0.05, \gamma = 0.5$  captures the experimental dynamics accurately within experimental uncertainties~\footnote{For example, for the data shown in Fig.~\ref{fig:fig4}a), a weighted goodness-of-fit analysis yields a reduced $\chi^2 = 3.15$ and a Pearson correlation coefficient of 0.996, confirming that the theoretical model reproduces the experimental trend accurately.
The somewhat elevated $\chi^2 $ value primarily reflects the very small experimental uncertainties ($\approx1\%$), which do not include systematic effects such as magnetic field jittering and amplitude calibration. When a realistic systematic uncertainty of a few percent is taken into account, the reduced  $\chi^2 $ approaches unity, indicating statistical consistency}, including the overall decrease of $|A(t)|$ as well as the superimposed damped 
oscillations (data for other values of $k_n a$ is shown in the Supplemental Material below).  
Furthermore, we find that parameter values in the ranges $\Gamma \simeq 0.05-0.1$ and $\gamma \simeq 0.5 - 2$ actually captures the  {\it both}
 the Cambridge~\cite{etrych2024universal} and the Aarhus experiments~\cite{morgenQuantumBeatSpectroscopy2023}. 
The fact that our phenomenological model can describe two independent experiments with consistent values of parameters indicates that it captures the most important  physics. 
The differences in the optimal values of $\gamma$ and $\Gamma$ for the two experiments can be partially explained by the differences in the trapping potentials
since  the value of $\Gamma_{pol}$ may be influenced by the in-homegeneity of the trap~\footnote{Note that we performed a trap averaging for ~\cite{morgenQuantumBeatSpectroscopy2023}, see the Supplemental Material below, and concluded that it has a weak effect on the time dynamics as the decoherence dynamics mainly happens at shorter timescales. It is known however that the presence of the trap broadens the polaron peak in the spectroscopic experiments~\cite{jorgensen2016}.}. 
Further, we account for magnetic decoherence and three-body losses only when analyzing the Aarhus experiment
 following the procedure described in Ref.~\cite{morgenQuantumBeatSpectroscopy2023}. Consequently, larger values of $\Gamma$ and $\gamma$ are required for the Cambridge experiment to accommodate these additional sources of broadening.
 
To enable the extraction of the experiment independent values of $\Gamma$ and $\gamma$, one would need to  exclude the broadening mechanisms specific to a given experimental setup. Such a systematic analysis is beyond the scope of the present work and is left for future studies.

{\it Summary.---} We investigated a phenomenological model of Bose polaron states, focusing on scenarios where the impurity is primarily correlated with a single boson, as these configurations couple most strongly to the two most commonly used experimental probes. To describe the possible decay of these  states 
to lower lying states that involve correlations with more bosons, we introduced an imaginary part to the impurity-boson 
interaction. Using a simple variational wave function, we then showed that this model can describe two recent measurements of
the spectral and dynamical properties of the Bose polaron, including both the energies and the decay rates. 

Our findings pave the way for deeper investigations into the limits of the quasi-particle description in the Bose polaron problem. First, our phenomenological approach offers a practical framework for parametrizing experimental data, thereby enabling meaningful comparisons across different experiments. 
Second, our findings underscore the need for microscopic theories of Bose polarons that 
include  the decay of the experimentally relevant polaron states. This entails analyzing the complex problem of few-body correlations and bound states within a many-body environment, which we expect to be most pronounced for light impurities. It would also be interesting to apply our model to 
systems with different masses and measurement protocols such as ejection spectroscopy~\cite{hu2016bose,Yan2020}.
Finally, the effects of a non-zero temperature~\cite{Knap2012,Schmidt_2018,dzsotjanDynamicalVariationalApproach2020,Guenther2018,Pastukhov2018,Field2020,Isaule2022} as well as of a lattice~\cite{Grusdt2014,Colussi2023,SantiagoGarca2024,alhyder2025} on the decay of the Bose polaron should be explored.

\acknowledgments{
We thank Georgios Koutentakis, Frédéric Chevy, Hussam Al Daas, and Richard Schmidt for fruitful discussions; Jan Arlt for sharing their experimental data and many fruitful discussions; Christoph Eigen for sharing their experimental data and inspiring discussions.
R.\ A., T.\ P. and G.\ M.\ B.\ have been supported in part by the Danish National Research Foundation through the Center of Excellence "CCQ" (Grant agreement no.: DNRF156), the Independent Research Fund Denmark- Natural Sciences via Grant No. DFF -8021-00233B.  R. A., A. V. and M. L. acknowledge support by the European Research Council (ERC) Starting Grant No.~801770 (ANGULON). R.\ A. received funding from the Austrian
Academy of Science ÖAW grant No. PR1029OEAW03.
}

Data availability. The data that support the findings of this
article are openly available~\cite{data}.

\bibliography{libraryUsed.bib}

\clearpage
\onecolumngrid
\setcounter{equation}{0}
\setcounter{figure}{0}
\setcounter{table}{0}
\setcounter{subsection}{0}
\renewcommand\theequation{S\arabic{equation}}
\renewcommand\thefigure{S\arabic{figure}}
\renewcommand\theHequation{S\arabic{equation}}
\renewcommand\theHfigure{S\arabic{figure}}

\begin{center}
\textbf{\large Supplemental Material for}\\[0.5em]
\textbf{Phenomenological model of decaying Bose polarons}
\end{center}

\subsection{Single excitation ansatz}
The Bogoliubov transformation is performed on the single channel Hamiltonian with its interaction term in Eq. (3) of the main text by moving to a basis of phononic excitations of the condensate. We write the boson operators as: $\hat{b}_{\mathbf{k}} = u_{\mathbf{k}}\hat{\beta}_{\mathbf{k}}-v_{\mathbf{k}}\hat{\beta}^\dagger_{-\mathbf{k}}$ with 
$
u_{\mathbf{k}} = \sqrt{(\epsilon_{\mathbf k}+n_B g_B)/(2E_{\mathbf k})+1/2}, 
    v_{\mathbf{k}} = \sqrt{(\epsilon_{\mathbf k}+n_B g_B)/(2E_{\mathbf k})-1/2},$  and   $ g_B=4\pi a_B/m. $
This gives the following Hamiltonian:
\begin{align}
\begin{split}
	\hat{H} &= \sum_{\mathbf{k}} 
	(\epsilon_{\mathbf{k}}+E_{\text{mf}}) 
	\hat{c}^\dagger_{\mathbf{k}} 
	\hat{c}_{\mathbf{k}}
	+\sum_{\mathbf{k}} 
	\xi_{\mathbf{k}} 
	\hat{\beta}^\dagger_{\mathbf{k}} 
	\hat{\beta}_{\mathbf{k}}
	+ \frac{\sqrt{N_B}}{\mathcal{V}}
	\sum_{\mathbf{q}\mathbf{p}} U_1(p)
	 \hat{c}^\dagger_{\mathbf{q}-\mathbf{p}} \hat{c}_{\mathbf{q}} 
	 (\hat{\beta}^\dagger_{\mathbf{p}} +\hat{\beta}_{-\mathbf{p}})
 \\&-\frac{1}{\mathcal{V}}
	 \sum_{\mathbf{k}\mathbf{q}\mathbf{p}}
	 U_2(p,k)
	 \hat{c}^\dagger_{\mathbf{q}-\mathbf{p}} \hat{c}_{\mathbf{q}+\mathbf{k}} 
	 (\hat{\beta}^\dagger_{\mathbf{k}}\hat{\beta}^\dagger_{\mathbf{p}} +\hat{\beta}_{-\mathbf{k}}\hat{\beta}_{-\mathbf{p}} )
	+\frac{1}{\mathcal{V}}
	 \sum_{\mathbf{k}\mathbf{q}\mathbf{p}}
	 U_3(p,k)
	 \hat{c}^\dagger_{\mathbf{q}-\mathbf{p}} \hat{c}_{\mathbf{q}-\mathbf{k}} 
	 \hat{\beta}^\dagger_{\mathbf{p}}\hat{\beta}_{\mathbf{k}}
	 \label{eq:Hamiltonian2}
\end{split}
\end{align}
where
$
E_{\text{mf}} = n_B g_I + \frac{g_I}{V}\sum_{\mathbf{k}}v_{\mathbf{k}}^2$ is the mean field energy. $ \xi_{\mathbf{k}} = \sqrt{\epsilon_{\mathbf{k}}(\epsilon_{\mathbf{k}}+2n_Bg_B)}$ is the phononic energy. $ U_1(p) = g_I (u_{\mathbf{p}} - v_{\mathbf{p}}) = g_I\sqrt{\epsilon_{\mathbf{p}}/E_{\mathbf{p}}}$ is the Fr\"ohlich interaction potential,
$U_2(p,k)= g_I u_{\mathbf{p}}v_{\mathbf{k}}$, and $U_3(p,k)= g_I (u_{\mathbf{p}}u_{\mathbf{k}}+v_{\mathbf{p}}v_{\mathbf{k}})$.
Here, we can relate $g_I$ to the boson-impurity scattering length using the Lippmann-Schwinger's equation:
\begin{equation}
g_I^{-1} = \frac{m}{4\pi a} - \frac{1}{V}\sum_{\mathbf{k}}^\Lambda \frac{1}{\epsilon^r_{\mathbf{k}}}
\end{equation}
where $\epsilon^r_{\mathbf{k}} = k^2/2m_r$ is the reduced energy of the impurity-boson system and $m_r = m m_I/(m+m_I)$ is the reduced mass. We note that $U_1(p) $, $U_2(p,k)$ and $U_3(p,k)$ only depend on the amplitudes of their arguments.
The goal is to study the evolution of the impurity dynamics considering one impurity is already present in the system and then at time $t=0$ we apply a `quench' to the system represented by a radio frequency which is considered weak enough for linear response to be valid. This quench creates one impurity state which gets dressed by excitation to form a polaronic cloud, and we are interested in the dynamics of the formation of the polaron state while including the effect of losses systematically.

We consider the following ansatz for the impurity state:
\begin{align}
	\begin{split}
	\ket{\Psi_1(t)} &= \Big( A \hat{c}^\dagger_0  + \sum_{\mathbf{k}} B({\mathbf{k}}) \hat{c}^\dagger_{-\mathbf{k}} \hat{\beta}^\dagger_{\mathbf{k}} \Big) \ket{\text{BEC}}
	\end{split}
\end{align}
that leads to the following equations of motion:
\begin{align}
    \begin{split}
&i\partial_t A = 
E_{\text{mf}}A
+
\frac{\sqrt{N_B}}{V}
\sum_{\mathbf{k}}
U_1(\mathbf{k})
B(\mathbf{k})
\\
&i\partial_t B(\mathbf{k})
=(\epsilon_{\mathbf{k}}+E_{\text{mf}}+\xi_{\mathbf{k}} )
B(\mathbf{k})
+\frac{\sqrt{N_B}}{V}
U_1(\mathbf{k}) A 
+\frac{1}{V}\sum_{\mathbf{p}}
U_3(\mathbf{k},\mathbf{p}) B(\mathbf{p})
\label{eq:4and5}
\end{split}
\end{align}

\subsection{Extended variational ansatz}
Here, we extend the previous ansatz with one more bath excitation. For that we employ the two-channel Hamiltonian in Eq. (1) of the main text which we can write in the Bogoliubov approximation as follows:
\begin{align}
\begin{split}
	\hat{H} &= \sum_{\mathbf{k}} 
	\epsilon_{\mathbf{k}} 
	\hat{c}^\dagger_{\mathbf{k}} 
	\hat{c}_{\mathbf{k}}
	+\sum_{\mathbf{k}} 
	\xi_{\mathbf{k}} 
	\hat{\beta}^\dagger_{\mathbf{k}} 
	\hat{\beta}_{\mathbf{k}}
	+\sum_{\mathbf{k}} 
	(\varepsilon^d_{\mathbf{k}} + \nu_0)
	\hat{d}^\dagger_{\mathbf{k}} 
	\hat{d}_{\mathbf{k}}
	+ g\sqrt{n_B}
	\sum_{\mathbf{k}}
	 (\hat{d}^\dagger_{\mathbf{k}} \hat{c}_{\mathbf{k}} 
	 +\hat{c}^\dagger_{\mathbf{k}} \hat{d}_{\mathbf{k}} )
 \\&+g \sum_{\mathbf{k}, \mathbf{q}}
 \Big[\hat{d}^\dagger_{\mathbf{q}} \hat{c}_{\mathbf{q}-\mathbf{k}} (u_{\mathbf{k}}\hat{\beta}_{\mathbf{k}}-v_{\mathbf{k}}\hat{\beta}^\dagger_{-\mathbf{k}})
	 \quad \quad \, \, \,   +
	 \hat{d}_{\mathbf{q}} \hat{c}^\dagger_{\mathbf{q}-\mathbf{k}} (u_{\mathbf{k}}\hat{\beta}^\dagger_{\mathbf{k}}-v_{\mathbf{k}}\hat{\beta}_{-\mathbf{k}}) \Big]
     \raisetag{2cm}
     \end{split}
\end{align}

We regularise the UV divergence by introducing a momentum cutoff $\Lambda$ and relating the molecular binding energy to the scattering length and an effective range $R_e$ using:
\begin{align}
	\begin{split}
		E_0 &= - \frac{g^2}{4\pi a} + \frac{g^2}{2\pi^2}\Lambda, \quad R_e = \frac{4\pi}{g^2}
	\end{split}
\end{align}
We write the following ansatz for the impurity state:
\begin{align}
	\begin{split}
	&\ket{\Psi_2(t)} = \Big( A \hat{c}^\dagger_0  + \sum_{\mathbf{k}} B({\mathbf{k}}) \hat{c}^\dagger_{-\mathbf{k}} \hat{\beta}^\dagger_{\mathbf{k}} + D \hat{d}^\dagger_{0} +  + \frac{1}{2} \sum_{\mathbf{k}\mathbf{p}} C({\mathbf{k}},{\mathbf{p}}) \hat{c}^\dagger_{-\mathbf{k}- \mathbf{p}} \hat{\beta}^\dagger_{\mathbf{k}} \hat{\beta}^\dagger_{\mathbf{p}} 
	+\sum_{\mathbf{k}} E({\mathbf{k}}) \hat{d}^\dagger_{-\mathbf{k}} \hat{\beta}^\dagger_{\mathbf{k}} 
	\Big) \ket{\text{BEC}}
    \raisetag{1.7cm}
	\end{split}
\end{align}
where $\ket{\text{BEC}}$ is the ground state of the system. We note that the impurity state is normalized to one. The equations of motion for the impurity state read as:
\begin{align}
	\begin{split}
		&i \partial_t A = g \sqrt{n_B} D - g \sum_{\mathbf{k}} v_{\mathbf{k}} E({\mathbf{k}})
		\\
		&i \partial_t B(\mathbf{k}) =  (\epsilon_{\mathbf{k}} 
		+ \xi_{\mathbf{k}} ) B(\mathbf{k}) + g u_{\mathbf{k}} D + g \sqrt{n_B} E({\mathbf{k}}) 
		\\
		&i \partial_t C(\mathbf{k},\mathbf{p}) = g\Big( u_{\mathbf{k}} E(\mathbf{p}) + u_{\mathbf{p}} E(\mathbf{k}) \Big)
		\\
		&i \partial_t D = E_0 D + g \sqrt{n_B} A + g \sum_{\mathbf{k}} u_{\mathbf{k}} B({\mathbf{k}})
		\\
		&i \partial_t E(\mathbf{k}) = \Big( \epsilon^D_{\mathbf{k}} + E_0 + \xi_{\mathbf{k}} \Big) E(\mathbf{k}) - g v_{\mathbf{k}} A + g \sqrt{n_B} B({\mathbf{k}}) + g \sum_{\mathbf{p}} u_{\mathbf{p}} C(\mathbf{k},\mathbf{p} )
	\end{split}
    \label{eq:extendedeom}
	\raisetag{1.5\baselineskip}
\end{align}

\subsection{Computational Method for the Eigenvalue Problem}
Gathering the variational amplitudes into the column vector  
\(\boldsymbol{\Psi}(t)=\bigl(A(t),\,B(\mathbf{k}_1,t),\,B(\mathbf{k}_2,t),\ldots\bigr)^{\mathrm T}\)  
reduces Eqs.~\eqref{eq:4and5} and \eqref{eq:extendedeom} to the concise Schr\"odinger-type form
\begin{equation}
   i\hbar\,\partial_t\boldsymbol{\Psi}(t)=H\,\boldsymbol{\Psi}(t),
\end{equation}
whose formal solution reads
\begin{equation}
   \boldsymbol{\Psi}(t)=\exp(-iHt/\hbar)\,\boldsymbol{\Psi}(0).
\end{equation}
which we can use directly to study the dynamics of the single excitation variational ansatz.

However, for the equations of motion in \eqref{eq:extendedeom}, the resulting dimension of the Hamiltonian matrix makes the use of the standard diagonalization routines computationally difficult. To circumvent this issue, the extended equations of motion can be recast as a generalised eigenproblem  
\begin{equation}
   \omega\,S\,|\Psi_2\rangle = H\,|\Psi_2\rangle ,
\end{equation}
where the diagonal matrix \(S\) contains the integration weights.  
Because \(S\) is positive definite we factorise it as \(S=T^{2}\) and introduce the transformed vector \(|\Phi\rangle=T|\Psi_2\rangle\).  This converts the problem to the ordinary Hermitian form  
\begin{equation}
   \omega\,|\Phi\rangle = \tilde H\,|\Phi\rangle, 
   \qquad \tilde H = T^{-1} H T^{-1},
   \label{eq:s14}
\end{equation}
which we diagonalise with an implicitly--restarted Lanczos routine~\cite{Yousef2003}.  For the parameters used in the paper (\(\Lambda=30,\;N_L=400, N_\theta = 10\)), the matrix \(\tilde H\) is of order $~10^5$ but remains sparse, allowing us to extract the lowest \(200\) eigenpairs efficiently. These are sufficient to resolve the part of the spectrum that governs the quench dynamics discussed in the main text.  Increasing slightly \(\Lambda\), \(N_L\), or the number of computed eigenvalues does not change qualitatively the results. 

To follow the quench dynamics we do not rely on the full spectrum but instead evaluate the action of the exponential \(\exp(-i\tilde H t)\) on the initial state by means of a short-iteration Arnoldi (Krylov) scheme.  
At each time step \(\Delta t\) we build an \(m\)-dimensional Krylov basis \(K_m(\tilde H,\Phi(t))\), compute the reduced Hessenberg matrix \(H_m\), and update the state via   
\[
   \Phi(t+\Delta t) \simeq \| \Phi(t) \|\,
   Q_m \exp(-i H_m \Delta t)\,e_1 ,
\]
where \(Q_m\) contains the orthonormal basis vectors and \(e_1\) is the first Cartesian unit vector.  
With \(m \approx 100\) and \(\Delta t\,/t_n=0.01\) this procedure converges to machine precision.
Physical observables are reconstructed from \(\Phi(t)\) by the inverse transformation \(|\Psi_2(t)\rangle = T^{-1}\Phi(t)\).
\subsection{Additional interaction values for dynamics}
\begin{figure}
	\centering
\includegraphics[width=0.5\columnwidth ]{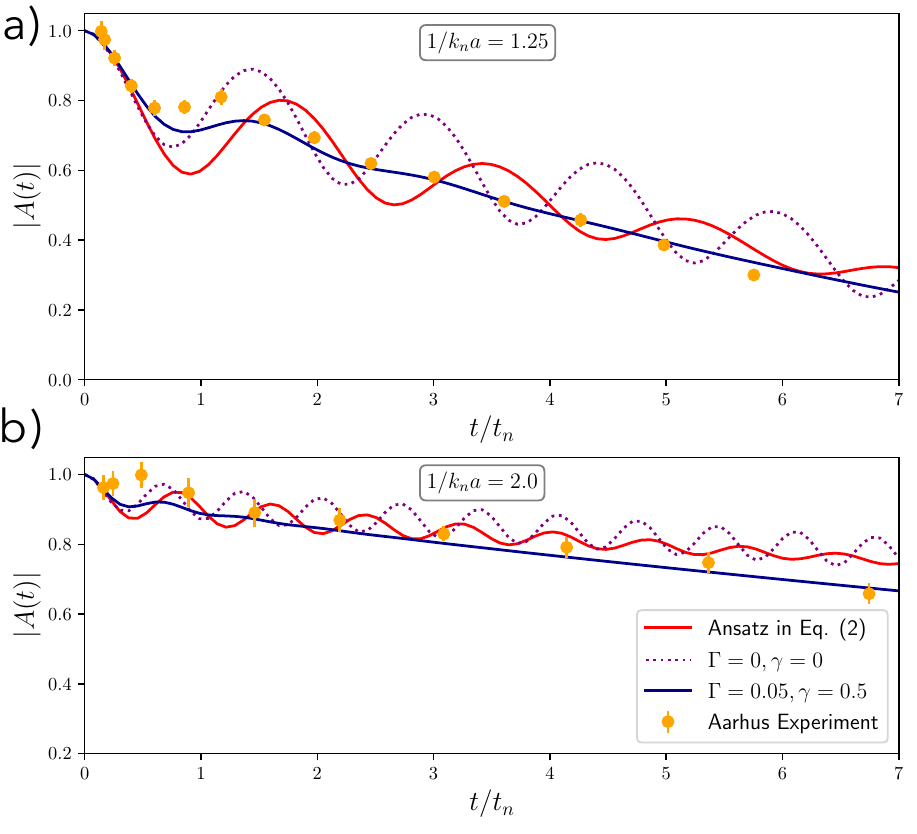}
	\caption[short]
{Time dynamics of the contrast $|A(t)|$ on the repulsive side of the Feshbach resonance for $(k_na)^{-1} = 1.25$ in panel $(a)$ and $(k_na)^{-1} = 2.0$ in $(b)$. Same format and parameters used in Fig. 3 of the main text.}
 \label{fig:figS4}
\end{figure}

Here, we present a comparison of our phenomenological ansatz with experimental data (see Fig.~\ref{fig:figS4}) at intermediate interaction strengths. Similar to the case of stronger interactions discussed in the main text, we find that incorporating decay through the complex parameter $g_I$ enhances the description of the dynamics. 

\subsection{Imaginary interaction term from three-body recombination}
We consider a system of bosonic particles and an impurity with the same mass. The scattering length between bosons is denoted by $a_{BB}$. The impurity-boson scattering length is denoted by $a$. We assume that $a$ is small. We also take into account three-body losses between the impurities and the bosons as a source of a non-unitary dynamics in the system. These losses occur at a rate given by $\Gamma_{3}$.

\begin{figure}
	\centering
\includegraphics[width=0.5\columnwidth]{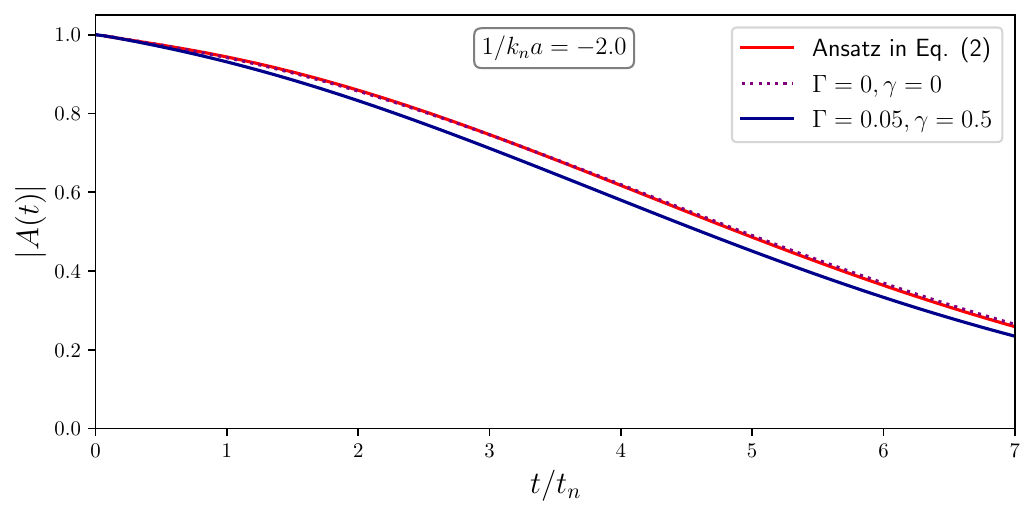}
	\caption[short]{
Time dynamics of the contrast $|A(t)|$ on the repulsive side of the Feshbach resonance for $(k_na)^{-1} = -2.0.$}
 \label{fig:all0}
\end{figure}

The Hamiltonian of such a system can be written as:
\begin{align}
	\begin{split}
		\hat{H} &= \hat{H}_{B} + \hat{H}_{I} + \hat{H}_{IB} + \hat{H}_{BB} + \hat{H}_{\text{loss}}
        \label{eq:Hamiltonian1}
	\end{split}
	\end{align}
with
\begin{align}
	\begin{split}
    &\hat{H}_{B} = \sum_{\mathbf{k}} \epsilon_{\mathbf{k}} \hat{b}^\dagger_{\mathbf{k}} \hat{b}_{\mathbf{k}}, \quad
    \hat{H}_{I} = \sum_{\mathbf{k}} \epsilon_{\mathbf{k},I}\hat{c}^\dagger_{\mathbf{k}} \hat{c}_{\mathbf{k}}n \quad \hat{H}_{BB} = 
	\frac{g_B}{2\mathcal{V}}\sum_{\mathbf{k},\mathbf{p},\mathbf{q}} \hat{b}^\dagger_{\mathbf{k}+\mathbf{p}} \hat{b}^\dagger_{\mathbf{q}-\mathbf{p}} \hat{b}_{\mathbf{k}} \hat{b}_{\mathbf{q}},\\
    &\hat{H}_{IB} =
	 \frac{g_0}{\mathcal{V}}\sum_{\mathbf{k},\mathbf{p},\mathbf{q}}\hat{b}^\dagger_{\mathbf{k}+\mathbf{p}} \hat{b}_{\mathbf{k}}\hat{c}^\dagger_{\mathbf{q}-\mathbf{p}} \hat{c}_{\mathbf{q}},
     \label{eq:hamIB}
     , \quad \hat{H}_{\text{loss}} = -i \frac{\Gamma_{3}}{2\mathcal{V}^{2}}
\sum_{\substack{\mathbf{k}_{1},\mathbf{k}_{2},\mathbf{q}_{1}\\
                \mathbf{q}_{2},\mathbf{q}_{3}}}
\,
\hat{c}^{\dagger}_{\mathbf{k}_{1}}\hat{c}_{\mathbf{k}_{2}}\,
\hat{b}^{\dagger}_{\mathbf{q}_{1}}\hat{b}_{\mathbf{q}_{2}}\,
\hat{b}^{\dagger}_{\mathbf{q}_{3}}\hat{b}_{\mathbf{k}_{1}+\mathbf{q}_{1}+\mathbf{q}_{3}-\mathbf{k}_{2}-\mathbf{q}_{2}},
	\end{split}
\end{align}
where $\epsilon_{\mathbf{k}} = \hbar^2 k^2/2m$, $\epsilon_{I,\mathbf{k}} = \hbar^2 k^2/2m_I$, $g_B$ is the interaction strength between two bosons and $g_0$ between an impurity and a boson.

Assuming the majority of bosons are condensed ($\hat{b}_0 = \hat{b}^\dagger_0 \simeq \sqrt{N_B}$) we can diagonalize the Hamiltonian by means of the Bogoliubov transformation. If we write the loss term to second order in the boson operators we can write:
\begin{align}
     \hat{H}_{\text{loss}} &\simeq -i \frac{2\Gamma_{3} n_B}{\mathcal{V}} \sum_{\mathbf{k},\mathbf{p},\mathbf{q}}\hat{b}^\dagger_{\mathbf{k}+\mathbf{p}} \hat{b}_{\mathbf{k}}\hat{c}^\dagger_{\mathbf{q}-\mathbf{p}} \hat{c}_{\mathbf{q}}
\end{align}
which we can add to the term in \eqref{eq:hamIB} modifying the interaction strength to $g_I = g_0 -2i\Gamma_{3} n_B $.

This construction illustrates one possible way to incorporate an imaginary interaction via the measured three-body loss rate $\Gamma_{3}$, other ways exist in the literature to account for decoherence effect that cannot be accounted for within a microscopic theory \cite{wangComplexContactInteraction2022}. However, we observed that the experimental value of $\Gamma_{3}$ \cite{skouNonequilibriumQuantumDynamics2021a} is too small to account for the observed decoherence.

The imaginary part used in the main text includes in principle all loss channels in the problem, including three-body losses and other broadening effects. For instance, the coupling to low lying states, the dimer state and scattering states related to them.

\subsection{Agreement in the attractive regime.}
For negative scattering length ($a<0$) all theoretical descriptions implemented in this work converge.  Since the impurity cannot form a bound dimer on the attractive side, higher-order scattering channels that differentiate the extended two-channel variational ansatz and the time-dependent treatment from the simple Chevy (single-excitation) ansatz are strongly suppressed.  Consequently, the predicted polaron coherence and energy shift are indistinguishable within experimental accuracy, as illustrated in Fig.~\ref{fig:all0}, where the three theoretical curves reproduce the experimental data for $1/k_n a=-2$~\cite{skouNonequilibriumQuantumDynamics2021a}.

\subsection{Trap averaging}
In the case of a harmonic trapping potential we can use the Thomas-Fermi approximation to write the density of the condensate as:
\begin{align}
	\begin{split}
	n_B(\textbf{r}) &= \frac{\mu - V(\mathbf{r})}{g_B}
	\end{split}
\end{align}
where $V(\mathbf{r}) = m/2\sum_i \omega_i^2 r_i^2$ is the trapping potential, $R_i = \sqrt{2\mu/m\omega_i^2}$ is the Thomas-Fermi radius and $\mu$ is the chemical potential. 
We integrate the equation:
\begin{align}
	\begin{split}
	N_B  = \int \, \text{d}\textbf{r} 
	\,
	n_B(\textbf{r}) 
	&= \left(\frac{2\mu}{\hbar\bar{\omega}^2}\right)^{5/2} \dfrac{a_\text{HO}}{15a}
	\end{split}
\end{align}
where $a_\text{HO} = \sqrt{\hbar/m\omega}$ is the harmonic oscillator length, $\bar{\omega} = (\omega_x\omega_y\omega_z)^{1/3}$ is the geometric mean of the trapping frequencies.
We can write the density of the condensate as:
\begin{align}
	\begin{split}
	n_B (r) &= \dfrac{15}{8\pi} \dfrac{N_B}{R^3} \left( 1 - \dfrac{r^2}{R^2} \right)
	\end{split}
\end{align}
The average condensate density $n_0 = \dfrac{N_B}{R^3} = 0.7 \times 10^{20}m^{-3}$. 
\begin{figure}
	\centering
\includegraphics[width=0.5\columnwidth]{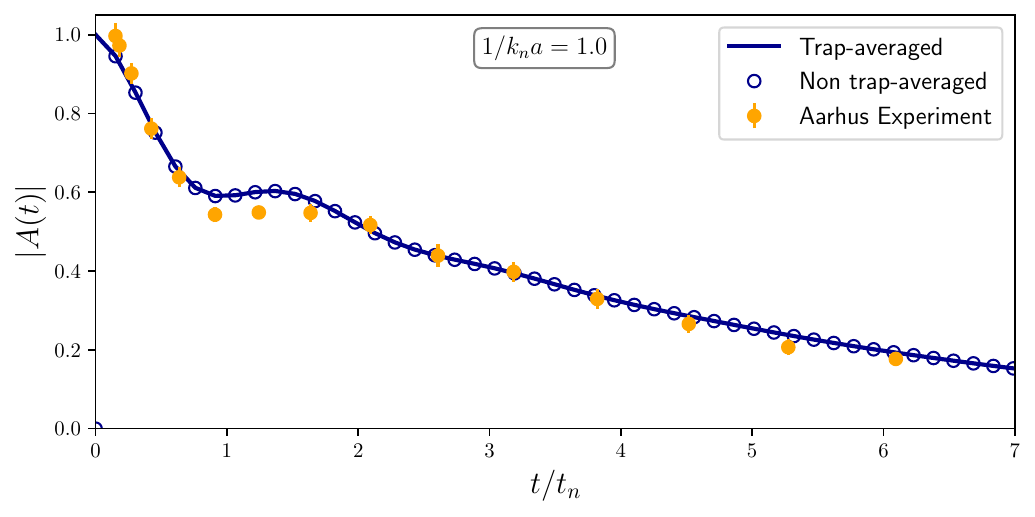}
	\caption[short]{
Trap averaged coherence function for the Aarhus experiment. In solid blue we show the result of our calculation, while in orange we show the experimental data.}
 \label{fig:trapavg}
\end{figure}
In our units, the coupled equations for the amplitudes $A$ and $B(k)$ in ~\eqref{eq:4and5} include $E_{\text{mf}}= n_B(r)g_I + \frac{g_I}{2\pi^2}\int dk k^2 v_{\mathbf{k}}^2$, $u_{\mathbf{k}} = \sqrt{(\epsilon_{\mathbf k}+4\pi n_B(r)(n_0^{1/3}a_B))/(2E_{\mathbf k})+1/2}$, 
, $\xi_{\mathbf{k}} = \sqrt{\epsilon_{\mathbf{k}}[\epsilon_{\mathbf{k}} + 8 \pi (n_0^{1/3}a_B) n(r)]}$, 
$W_1(k) = 2  g_I \sqrt{\frac{\epsilon_{\mathbf{k}}}{ 2\pi \xi_{\mathbf{k}} }}$, $W_3(k,p) =  \frac{g_I}{2\pi^2}(u_{\mathbf{p}}u_{\mathbf{k}}+v_{\mathbf{p}}v_{\mathbf{k}})$, with $n(r) = \dfrac{15}{8\pi} \left( 1 - \dfrac{r^2}{R^2} \right)$. 
We plot in Fig. \ref{fig:trapavg} the coherence function for one value of $1/k_n a$ with and without trap averaging. We can see that the coherence function is not affected by the trap averaging using the same values of $\Gamma, \gamma$ as the main text.

\end{document}